# Eficiency of REST and gRPC realizing communication tasks in microservice-based ecosystems

Marek BOLANOWSKI[a,1], Kamil ŻAK, Andrzej PASZKIEWICZ[a], Maria GANZHA[b],
Marcin PAPRZYCKI[b], Piotr SOWIŃSKI[b], Ignacio LACALLE[c], Carlos E. PALAU[c]

[a] *Rzeszow University of Technology, Rzeszów, Poland*
[b] *Polish Academy of Sciences, Warszawa, Poland*
[c] *Communications Department Universitat Politècnica de València, Valencia, Spain*

ORCiD ID: Marek Bolanowski https://orcid.org/0000-0003-4645-967X, Kamil Żak https://orcid.org/0000-0001-7497-8608, Andrzej Paszkiewicz https://orcid.org/0000-0001-7573-3856, Maria Ganzha https://orcid.org/0000-0001-7714-4844, Marcin Paprzycki https://orcid.org/0000-0002-8069-2152, Piotr Sowiński https://orcid.org/0000-0002-2543-9461, Ignacio Lacalle https://orcid.org/0000-0002-6002-4050, Carlos E. Palau https://orcid.org/0000-0002-3795-5404

**Abstract.** The aim of this contribution is to analyse practical aspects of the use of REST APIs and gRPC to realize communication tasks in applications in microservice-based ecosystems. On the basis of performed experiments, classes of communication tasks, for which given technology performs data transfer more efficiently, have been established. This, in turn, allows formulation of criteria for the selection of appropriate communication methods for communication tasks to be performed in an application using microservices-based architecture.

**Keywords.** REST, gRPC, .NET, distributed systems, IoT, microservices, adaptive networks.

## 1. Introduction

The main impulse behind data transfer research is the optimisation of communication between components in a distributed system, with particular emphasis placed on microservice architectures [1,2]. In such architecture, performance of the implemented system is directly influenced by the choice of communication mechanisms between individual microservices. Note that, today, microservices have become more and more popular, and find their way to large-scale heterogeneous deployments, such as in the Internet of Things (IoT), Industry 4.0 [3-7] including cyber-physical systems [8-9]. Dedicated methodologies have also been developed for this class of solutions [10]. During preliminary research, the popularity of various communication technologies that can be used in this context was analysed using, among others, Google Trends.

---

[1] Corresponding Author: Marek Bolanowski, Department of Complex Systems, Faculty of Electrical and Computer Engineering, Rzeszow University of Technology, al. Powstańców Warszawy 12, 35-959 Rzeszów, Poland; E-mail: marekb@prz.edu.pl.

Specifically, the level of interest in particular technologies has been reported in [11]. The conclusion was simple. The undisputed leader among web application interfaces, in the IT market, is the REST API [12-15]. The second, in terms of popularity, turned out to be the gRPC framework [16-17]. Currently, gRPC is less popular than REST API, but the growing interest in it, as evidenced by its inclusion in the CNCF (Cloud Native Computing Foundation) project [18], is worth noting.

While there exist multiple reasons for selecting either of the two solutions, it is worth to observe that practical performance, in large-scale distributed deployments (e.g. in IoT ecosystems), should directly influence the discussion. This is because the size (understood, for instance, in terms of the number of components that have to communicate, or the physical and logical arrangement of the virtual elements to connect) of actual IoT deployments will grow. This growth will be influenced not only by the size of individual IoT ecosystems, but also by the need to combine (join) them to deliver more advanced services to the users. Thus, the main objective of this work is to address the question: which of the two technologies performs data transfer more efficiently for a specific class of communication tasks? In this context, it was decided that the analysis will be carried out when service implementations are built using the .NET platform and the C# programming language [19]. Obviously, it is possible to claim that different results could have been obtained when different implementation decisions were made. However, this argument could be seen as an opportunity for other works to step in, and to complete similar analysis using different stacks and setups. According to the authors, this is much needed in the context of software design for large-scale, distributed, communication-driven ecosystems.

It is also worth noting that, at the initial stage of application design, or when planning integration of existing systems, analysts should be provided with guidelines as to what type of communication technologies should be used for specific microservices transmitting a given type of data. In highly interconnected systems, even a small reduction in communication delays between microservices can, through synergistic effects, significantly influence the responsiveness of the entire system. The results of the preliminary research work, which are presented in what follows, allow formulation of criteria for selecting REST and gRPC technologies depending on classes of communication tasks that are to dominate in the deployed system.

*1.1. Available comparisons and reports*

Currently, there are few studies that address (and compare) communication performance of REST API and gRPC. The book [1] presents the use of REST, gRPC, and other technologies for synchronous and non-synchronous communication, in the design of microservice architecture applications. Its authors identified REST and gRPC as the most widely used protocols for synchronous messaging, in line with the observations noted above. However, the comparison is limited only to a theoretical description, which points out the potential advantage of the gRPC framework, due to its support of the HTTP/2 protocol. Work reported in [20] compares battery consumption (on Android phones) using communication based on REST, SOAP, Socket, and gRPC. A similar approach can be found in [21], which concerns migrating complex computational processes from a phone to an external server, using different implementations of the gRPC framework, on Linux Debian and Android operating systems. Here, authors refer to the research conducted in [20]. While interesting, battery consumption, as communication performance indicator is given little attention in comparison to others. For instance, it is usually deemphasized when analysing communication in large-scale industrial IoT deployments. In a paper [22], the authors identified REST as one of the leading standards for communication between microservices, without examining its impact on the overall system performance.

Interesting, from the point of view of the conducted research, is an article available on the blog of the developers of the gRPC framework. It compares the performance of synchronous communication in client applications on Android [23]. In the experiments the speed of data transfer from an HTTP JSON application and a gRPC-based client was examined. It was focused on a separated comparison of the receiving and the sending phases of the process. In practice, this mainly boils down to a comparison of data serialization and deserialization performance for JSON and Protobuf (the information exchange mechanism of gRPC). In [24], the authors compared the performance of REST, gRPC and Apache Thrift, measured in terms of the load on individual elements of the NUMA machine, on which the client applications reside; where the application server was built following a microservice structure. The authors pointed out the advantages of gRPC and Thrift technologies over REST.

The available analyses focus mainly on the description of applications of the discussed technologies and the comparison of their performance in narrow scenarios and/or application areas. It can be also observed that they are often focused on mobile applications. Therefore, it has been decided to slightly broaden the spectrum of analyses, by using different scenarios, found in practical applications. The selected test scenarios are based on authors experiences in the industry and in EU-funded research. They represent most common classes of communication tasks, used in the design and implementation of microservice-based systems.

## 2. Methodology and experimental setup

To carry out the proposed explorations, implementations of REST and gRPC services were prepared. They were built using the .NET 5 platform, with the use of ready-made templates: NET Core Web API – generating code for a web application, compliant with the assumptions of the REST standard, using communication with HTTP JSON; NET Core gRPC Service – implementation of the gRPC framework, allowing to build web applications.

Originally, various types of software performance testing methods [25-28], including: load testing, stress testing, endurance testing, spike testing, volume testing, have been considered. Upon further analysis of the context (large-scale, distributed, message driven ecosystems, e.g. IoT deployments) a combination of several test types has been chosen. One of the main parameters checked, when testing communication characteristics of a system, is the response time. It consists of the time needed to (1) establish a connection, (2) send a request, (3) process the service logic, and (4) return a response. The tests were carried out using the Apache JMeter application [29-30], a testbed using Novus One Plus platform and the IxLoad application [31]. Two groups of devices (client and servers) were used to perform the tests to simulate the communication of microservices over an actual network connection. At this stage of the work, Network Emulator II [32] was used to monitor the interference on the communication path used by microservices located on individual machines. A schematic of the test stand is shown in Figure 1.

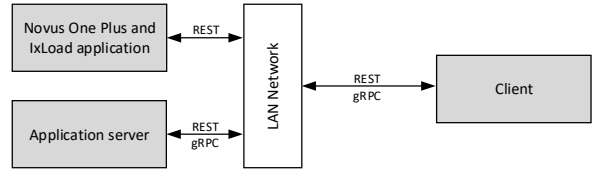

**Figure 1.** Test stand scheme.

The purpose of each test scenario was to measure the server response time, to a request sent by the test application. The structure of a sample request, along with an indication of the tested time range, is shown in Figure 2.

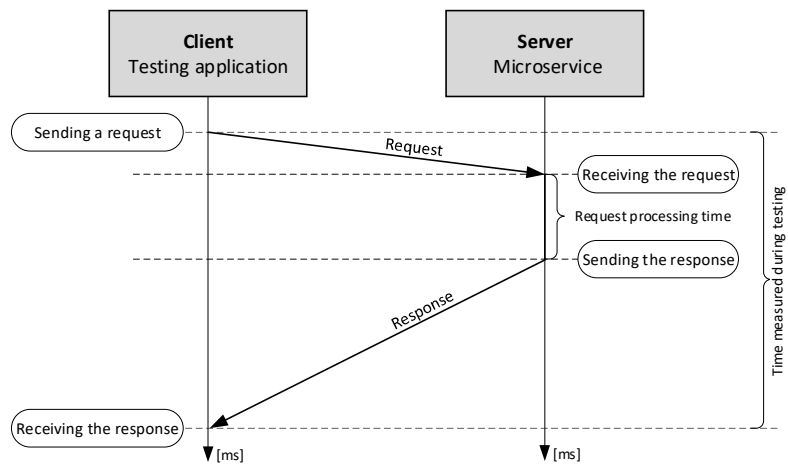

**Figure 2.** Diagram of a sample request detailing the range of time measured during the test.

Most of the measurements were performed using the JMeter application. However, additional validation measurements were performed, on an additional test stand, using the Ixia Novus One Plus system with the IxLoad application.

*2.1. Research scenarios*

Each test scenario studies the behaviour of a different type of services, depending on different operating parameters. The main factor tested during the study was the response time of the service (see above). It is the time measured from the moment of sending a request, from the testing application, to the end of receiving a response from the microservice (see Figure 2). To enrich the testing, the behaviour of the application has been emulated under different conditions, via setting different options (varying among tests). These parameters include: type of tested service method; number of sent requests per time unit; use of data encryption (TLS protocol). For the tests, two microservices, implemented as web services, based on the .NET platform were prepared. The first one is a standard REST application, using data exchange (in JSON format). The second application is a service using the gRPC framework, using data transport in a binary form. In each application, a set of methods was created. Once started, the implemented methods return data in response to incoming requests. When building the application, care was taken to ensure the diversity of supported data types. As a result of method calls, primitive data types, arrays, objects, or file data could be received. All requests have REST and gRPC counterparts. This means that methods, tested in a given scenario, have identical input data structure and return the same responses. The following communication tasks were used in the tests:

- Text cloning – takes as parameters any character string and a number that indicates how many times the provided string should be cloned in the response. Returns a cloned character string, depending on the given parameters. The size of the listed data is proportional to the input value.
- Get integer – parameterless method that returns the number 2147483647, which is the maximum number in the range of the Int32 type, in the .NET libraries. The size of the data exchanged during the request is small.
- Getting an array of consecutive integers – has one parameter, which is the number of elements of the returned array. It has a significant impact on the size of the information sent in the request.
- Fetch text file – a parameterless method that returns a binary string representing a text file of size equal to 455 KB. The file data is returned as a bit array.
- Download PDF file – a parameter-free method that returns a binary string representing a PDF file, of size equal to 3.4 MB. This request exchanges a relatively large amount of information. Data is transferred as a bit array.

For the sake of uniformity, each test performs a fixed number of requests to the application, over the course of five minutes, allowing for easy comparison of results from several scenarios. The performance of the services has been analysed for operations using information encryption – the TLS protocol. Each test scenario contains two timed runs, the first without the use of encryption (HTTP requests) and the second with its use (HTTPS requests). It is also worth mentioning that the tested applications do not use a database – the mechanism responsible for the returned responses was implemented inside the libraries. In standard (web-based) applications, database access usually accounts for a large part of the service response time. However, for communication performance analysis purpose this seems unnecessary, as both scenarios (REST and gRPC) have been designed without this element, removing any potential bias element related to database read/write operations. Additionally, the response time of identical queries to the database considerably varies between requests. However, this study focuses on analyzing the behavior of the data transport layer only. Therefore, the lack of a database, and other external factors, was an explicit decision aimed at minimizing the risk of a their non-negligible impact on the timing results.

## 3. Experimental results

In this section, results of the tests conducted for individual services under the assumption of a low load of requests are summarized. Specifically, it was assumed that, for this configuration, in each scenario, the service will send requests at the rate of one request per second (i.e. three hundred requests in five minutes, see above). Both the "without encryption" and "with encryption" variants have been tried.

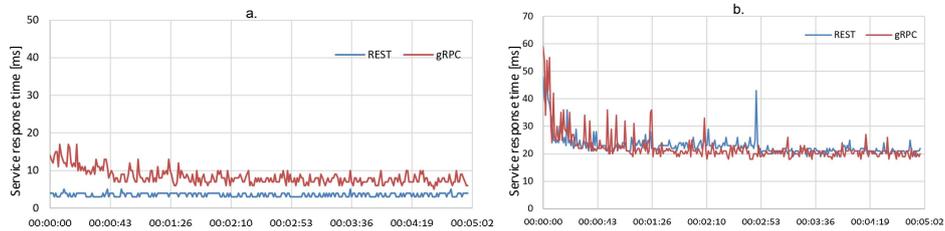

**Figure 3.** Service response times for test Scenario I: a. without encryption; b. with encryption.

The first test, Scenario I, focuses on comparing the response times generated by the services method, cloning text with low data overhead. The text "Hello World!" was passed as a parameter, in the form of a string of characters. The service returned an identical string as a response. The response times are shown in Figures 3a and 3b. There are discrete spikes in the values, which could have been caused by various factors, such as the way tasks are queued in the operating system, or the availability of computing power. For the first graph, the median response times was 4 ms for REST applications, and 8 ms for gRPC. Additionally, all requests sent to the REST applications were handled faster, compared to the requests sent to the gRPC applications. The situation is slightly different for the test with data encryption (3b). In this case, the medians are very similar, i.e. 22 ms and 21 ms for the REST and the gRPC services, respectively. The values obtained were higher and had slightly more fluctuations, especially in the initial phase of the test. In contrast to the previous test variant, the gRPC service obtained results similar to REST.

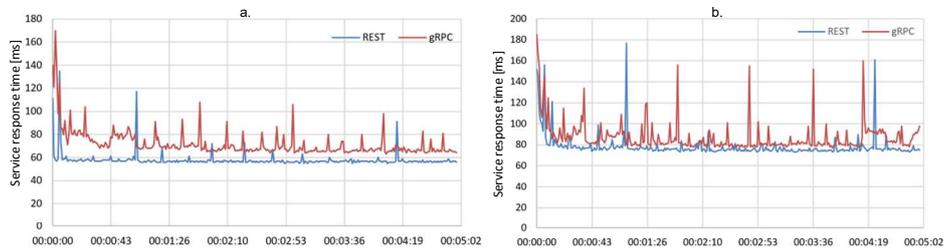

**Figure 4.** Service response times for test Scenario II: a. without encryption; b. with encryption.

In the next test, Scenario II, the same method was used, but it was parameterized in such a way that the received response has a larger size. One paragraph of the popular typeface presentation text "Lorem ipsum..." was sent as a parameter. Its length is 615 characters. To further increase the amount of information returned, a second parameter (number) was used to make the service clone the received text 1000 times and submit it as a response. Graphs showing the response rates of the services are presented in Figure 4. In both cases, the advantage of the REST service is evident, both for the case without data encryption (4a) and with its application (4b). In the first configuration, the median response time of the REST service was 57 ms, while it was 68 ms for the gRPC service. The configuration using HTTPS presents median results with values of 76 ms and 82 ms, for the REST and the gRPC applications, respectively. The speed advantage for REST was slightly offset in the second variant of the test, but it was still evident.

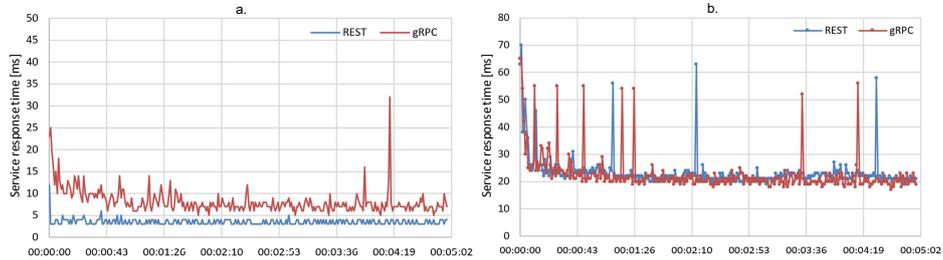

**Figure 5.** Service response times for test Scenario III: a. without encryption; b. with encryption.

For the next test (Scenario III), the methods that return integer value 2147483647 was used. This parameterless method generates answers with very low data overhead. The results are shown in Figures 5a and 5b. Here, it can be seen that the response times of both applications are very small. This is mainly due to the size of the data that had to be returned to the requester. Regarding the response times, in the first configuration, the median was 3 ms for the REST service, and 7 ms for the gRPC service. For the data encryption, it was 22 ms for the REST, and 21 ms for the gRPC. Thus, with requests of this type, REST performed better in the scheme without data encryption, while using HTTPS allowed gRPC to gain an almost imperceptible advantage in response speed.

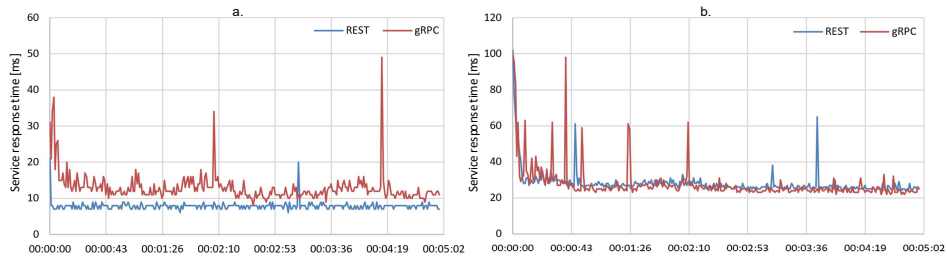

**Figure 6.** Service response times for test Scenario IV: a. without encryption; b. with encryption.

The next test, Scenario IV, involves sending requests to a method that returns a sequence of numbers from 0 to 10,000 in the form of an array. Compared to the previous case, the amount of data returned is significantly larger. The results are presented in Figures 6a and 6b. In the test scenario in the transmission without data encryption, once again, the response times of the REST service were faster with a median of 8 ms. Meanwhile, the gRPC application completed requests with a median value of 12 ms.

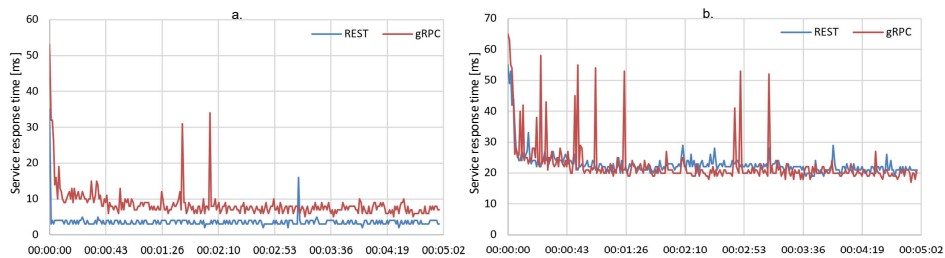

**Figure 7.** Service response times for test Scenario V: a. without encryption; b. with encryption.

When using the service versions in conjunction with HTTPS, the median time was 27 ms and 25 ms for the REST and the gRPC, respectively, indicating a minimal advantage for the framework using binary information transfer.

Microservices, in most cases, use representation of information in the form of objects. The next test (Scenario V) analyzes the process of retrieving data with such a structure. It uses a method that returns sample data of small size. The results are shown in Figures 7a and 7b. The medians of the results received for the first graph were 3 ms for the REST service, and 8 ms for the gRPC service. When data encryption was used, the median response times, of the REST application, were 22 ms, while for the gRPC application, the median was approximately 1 ms less. Note that in the second configuration, the gRPC framework showed more fluctuation in request response times, especially in the initial phase of the test. Similar to the previous test scenarios, the difference in service response speed is equalized in the case of HTTPS.

The next test, Scenario VI, was conducted to examine the behaviour of both server-side services (REST- and gRPC-based) when sending small-sized files to the client. A method that returns a text file in response was used. The test results are shown in Figures 8a and 8b. In this test scenario, it was the first time that the gRPC service showed an advantage in both configurations. Requests without data encryption generated results with a median of 57 ms for the REST application and 53 ms for the gRPC. The advantage was even more pronounced when using HTTPS, where the medians took values of 75 ms for the REST and 66 ms for the gRPC framework. This result may be related to the fact that the gRPC service does not need to convert the bitstream of the file, sending it to the client in the same form. The REST application interface bases its operation on the transport of information in the text form (JSON), forcing an additional conversion.

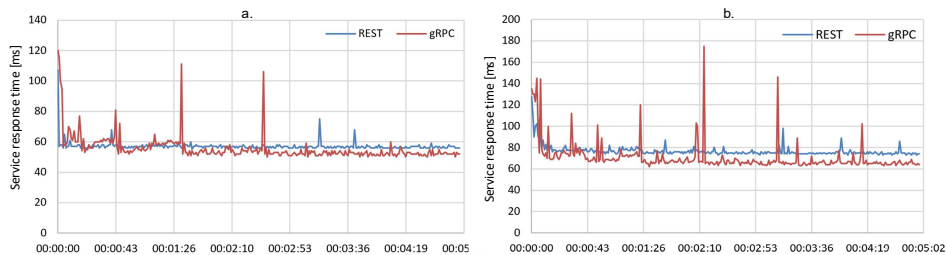

**Figure 8.** Service response times for test Scenario VI: a. without encryption; b. with encryption.

The next test, Scenario VII, was similar to the previous test (Scenario VI) and examined the performance of the services when sending a (slightly larger) file. For this purpose, a method was used that returns a sample PDF file. Application response times are presented in Figures 9a and 9b.

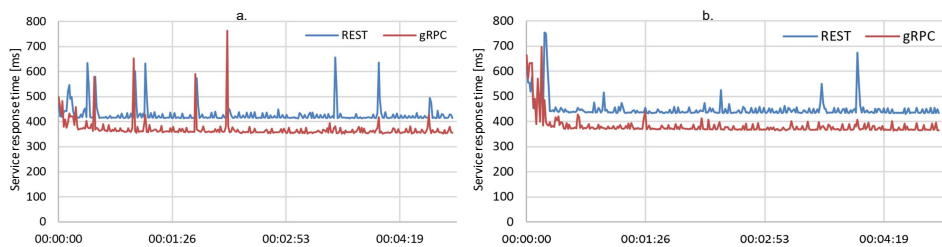

**Figure 9.** Service response times for test Scenario VII: a. without encryption; b. with encryption.

Use of the gRPC service, in transferring a file of significant size, also shows better performance compared to the REST. In the test presented in Figure 9a, the median value obtained was 416 ms for the application with the REST interface, while it was only 361 ms for the application based on the gRPC framework. The data encryption, presented in Figure 9b, shows median values of 473 ms and 370 ms for the REST and the gRPC, services, respectively. In both cases, the difference in the received response times is significant. This provides clear indication of the advantage of data transfers using the gRPC framework when dealing with larger binary-formatted files.

The architecture of the Ixia Novus One Plus traffic generator, and the IxLoad application, minimizes the latency contributed by the operating system, within which the client sending the requests resides. This software/hardware measurement system (which uses FPGAs), does not currently support tests for the gRPC technology. Validation tests are therefore limited to the REST technology. However, this approach allows to verify that fully software-based measurements (e.g. using JMeter) may be affected by significant measurement errors (latency contributed by the measurement application itself). In other words, the tests conducted in this scenario were set up to determine whether the software-based latency measurement architecture, used in Scenarios I through VII had any significant impact on the results. In this test, Scenario VIII, which is a mapping of Scenario VI, response times during text file retrieval were studied. The times were generated only for the REST service. Obtained results are shown in Figures 10a and 10b.

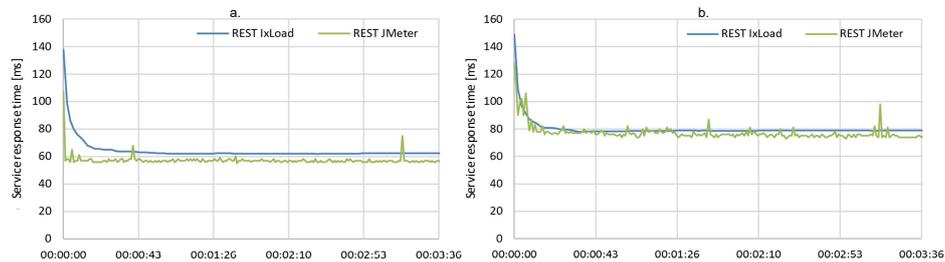

**Figure 9.** Service response times for test Scenario VIII: a. without encryption; b. with encryption.

The median REST application response time, measured using IxLoad, was 69.2 ms for the configuration without data encryption, while it was 74.82 ms when using encryption. These results are reasonably close to the results obtained in the Scenario VI. This demonstrates the reliability of reported results. It should be stressed that similar results were obtained for the other six scenarios. It is also noteworthy that the graphs obtained using IxLoad, do not have discrete value spikes, which appeared in the tests performed using the JMeter toolset.

*3.1. Additional analysis of results*

Meta-level analysis of obtained results allows one to develop the criteria for the selection of a given communication technology in the context of the implemented communication tasks in distributed, message-driven ecosystems. The resulting list of criteria, and results of their application, is presented in Table 1. Based on the analysis of the individual scenarios, the table indicates the areas, in which a given technology (REST or gRPC) is characterized by better communication efficiency. In the case of similar results in a given area, both technologies were marked as suitable for use by communicating microservices.

**Table 1.** Evaluation of REST and gRPC technologies in terms of their exploitation areas.

| Area examined | No encryption | | With encryption | |
|---|---|---|---|---|
| | REST | gRPC | REST | gRPC |
| Transmission of text data (Scenario I, Scenario II) | × | | × | × |
| Transmission of numerical data (Scenario III) | × | | × | |
| Transmission of numerical data – large data size (Scenario IV) | × | | × | × |
| Transmission of structured data in the form of tables and objects (Scenario V) | × | | × | × |
| File transfer – small-sized (Scenario VI) | × | × | | × |
| File transfer – large-sized (Scenario VII) | | × | | × |

The selection criteria shown in Table 1 can be used to choose the optimal operating conditions for both technologies with respect to given communication tasks and their implementation in .NET. The REST based interface can be used for most standard servers that transfer simple data of relatively small size, which is retrieved according to the current needs of the client. The legitimacy of the REST applications is also valid for the textual data. The use of the gRPC interface can also be justified in specific applications, for example, in systems where periodic requests to transfer relatively large files occur in addition to the standard exchange of messages between microservices. Another application, where the use of the gRPC is worth considering, is in systems where a high volume of data is "non-stop" exchanged between microservices. Performed experiment have also shown that when using data encryption, in standard applications, the gRPC framework gains a slight advantage over the REST, in terms of communication performance.

It should be stressed that, in the case of the need for continuous information exchange, when the number of messages transmitted between applications is very large, even small differences in the latency of individual exchanges can have very large impact on the performance of the distributed system as a whole. Such situation may occur relatively often in applications with microservices architecture and in (very-)large distributed IoT ecosystems. For this reason, the proper selection of inter-node communication technology can be very important.

## 4. Concluding remarks

This contribution analyses the performance of communication tasks realized using the two most popular technologies of message exchange in applications built on the basis of microservices, i.e. REST API and gRPC. At the initial stage of research, core communication tasks, encountered in applications built for research and commercial purposes were selected. For each such operation, performance tests, comparing the execution times of communication tasks were performed. On this basis, a set of initial recommendations has been developed, which can be used when building applications with a highly distributed architecture with heterogeneous inter-node query structure. This approach seems to be appropriate both in the case of IoT, and for Industry 4.0 applications. Departure from the homogeneous (realized exclusively by a single standard) structure of message handling in a given application, may significantly reduce the overall delay in the responsiveness of the system as a whole, especially whenever the number of messages exchanged between applications is expected to be large.

In further work, additional technologies will be evaluated such as: GraphQL, OData, WSDL, and the set of partial communication tasks will be extended. In addition, it may be worth to replicate this experiment with different stacks (not .NET services, other testing tools, varying message patterns and loads). This should allow extending the set of recommendations and, eventually, they may result in the development of an expert system that supports the design decision-making process during the development of application architecture.

**Acknowledgements:** This project is financed by the Minister of Education and Science of the Republic of Poland within the "Regional Initiative of Excellence" program for years 2019–2022. Project number 027/RID/2018/19, amount granted 11 999 900 PLN.
To carry out the research and to verify obtained results, the "Stand for research on phenomena in the environment of the Internet of Everything" was used, located in the Department of Complex Systems of the Rzeszow University of Technology [33].
Work of Maria GANZHA, Marcin PAPRZYCKI, Piotr SOWIŃSKI, Ignacio LACALLE, and Carlos E. PALAU, was completed within the scope of the ASSIST-IoT project that has received funding from the European Union's Horizon 2020 research and innovation program under grant agreement 957258.